\documentclass[usenatbib]{mn2e}
\usepackage{color}
\usepackage{float} 
\usepackage[fleqn]{amsmath}
\usepackage{epsfig,floatflt}
\usepackage{natbib}
\usepackage{subfigure}
\usepackage{amssymb}
\usepackage{multirow}
\usepackage{ulem}
\usepackage{bm}
\usepackage{url}
\usepackage{placeins}

\newcommand{\Mpch}{{$h^{-1}$Mpc}}
\newcommand{\kmsMpc}{{km.s$^{-1}$.Mpc$^{-1}$}}
\newcommand{\kms}{{km.s$^{-1}$}}
\newcommand{\LCDM}{$\Lambda$CDM}

\author[G. Lavaux]{G. Lavaux\\
Department of Physics, University of Illinois at Urbana-Champaign, 1110 W. Green St, Urbana, IL, 61801, USA}

\title{Precision constrained simulation of the Local Universe}

\begin{document}

\maketitle

\begin{abstract}
  We use the formalism of constrained Gaussian random fields to
  compute a precise large scale simulation of the 60\Mpch{} volume of
  our Local Universe. We derive the constraints from the
  reconstructed peculiar velocities of the 2MASS Redshift Survey. We
  obtain a correlation of $0.97$ between the log-density field of the
  dark matter distribution of the simulation and the log-density of
  observed galaxies of the Local Universe. We achieve a good
  comparison of the simulated velocity field to the observed velocity
  field obtained from the galaxy distances of the NBG-3k. At the end,
  we compare the two-point correlation function of both the 2MRS
  galaxies and of the dark matter particles of the simulation. We
  conclude that this method is a very promising technique for exploring
  the dynamics and structure of the Universe in our neighbourhood.
\end{abstract}

\section{Introduction}

We present a new method to re-simulate the large scale structures of
the Local Universe to a high precision. Re-simulating the dynamics of the large scale structures in our neighbourhood is an important step to understand the
environmental effects that may affect galaxy formation in our part of
the Universe and to have better constraints on the potential systematic
effects in the observations. Previous attempts, such as \cite{Kolatt96}
and \cite{Mathis02}, used a global constraint of the smoothed density
field, which is then transported back in time using
Bernoulli-Zel'dovich equation in a 80\Mpch{} volume. \cite{Nar01}
used PSCz and a perturbative expansion modelling of the evolution of the
density field to recover initial conditions in a 50\Mpch{}
volume. \cite{Fontanot03} used NOG and PSCz and a semi-Lagrangian
reconstruction procedure to obtain the reconstructed displacements in
a 60\Mpch{} volume. A different approach was followed by
\cite{Klypin03} and \cite{K02} who used Mark III peculiar velocity survey and applied
linear theory to obtain pure dynamical constraints, though rather
sparse and in a 30\Mpch{} volume. Related constrained simulations were then
performed in the same manner \citep[e.g][]{MV07} to study specific local environmental effects. All these reconstructions managed to capture the main large 
structures but the smaller scales ($\sim$\Mpch{}) were not well recovered. 

In this work, we use the Monge-Amp\`ere-Kantorovitch (MAK)
orbit reconstruction method \citep{Brenier2002,mohtu2005,Lavaux08} to recover a
dense set of constraints directly in Lagrangian coordinates. We use
the catalogue of the reconstructed orbits of the 2MASS Redshift Survey
\citep{TwoMRS,HuchraIAU05} already obtained in
\cite{LavauxTully09}. We put the constraints object by object using
a local adequate Lagrangian smoothing scale which depends on the
estimated mass used for the reconstruction procedure.

In Section~\ref{sec:method}, we present the details of the used
methods. In Section~\ref{sec:ic}, we specify the set up of initial
conditions from the 2MASS Galaxy Redshift survey. In
Section~\ref{sec:results} we discuss our results and in
Section~\ref{sec:conclu} we conclude.

\section{Methods}
\label{sec:method}

In this Section, we describe the general methods used to compute and apply the constraints on the initial conditions of the $N$-body simulation. In Section~\ref{sec:gcr}, we recall the theory of constrained Gaussian realisation. In Section~\ref{sec:mak}, we recall the fundamentals of the MAK reconstruction.

\subsection{Gaussian constrained realisation}
\label{sec:gcr}

To build the initial conditions of our simulation, we are using a
constrained Gaussian realisation algorithm \citep{BertWeygaert},
derived from the Hoffman-Ribak method \citep{HR91}. This
algorithm allows us to put any linear constraint of a Gaussian random
density field, which corresponds to the initial condition of an
$N$-body simulation. We give here a summary of the steps of the algorithm.
We first compute the variance matrix $V_{i,j}$ of the kernel of the constraints:
\begin{equation}
  V_{i,j} = \int \text{d}^3 {\bf k} P(k) \hat{K}^{*}_i({\bf k})\hat{K}_j({\bf k})
\end{equation}
with $K_i({\bf k})$ the kernel window function given in appendix F of \cite{BertWeygaert}.
We build the Fourier transform $\hat{\delta}_\text{r}({\bf k})$ of a Gaussian random field and then compute the constrained realisation
\begin{equation}
  \hat{\delta}({\bf k}) = \hat{\delta}_\text{r}({\bf k}) + \sum^{N_\text{c}}_{i,j=1} V^{-1}_{i,j} (c_j - c^\text{m}_j) K_i({\bf k})
\end{equation}
with $c_j$ the value of the constraint, $N_\text{c}$ the number of constraints and
\begin{equation}
  c^\text{m}_j = \int \text{d}^3{\bf k} K_j^{*}({\bf k}) \hat{\delta}_\text{r}({\bf k}).
\end{equation}
After the inverse Fourier transform, the field $\delta({\bf r})$
satisfies both the constraints and the power spectrum.  Practically all the
integrals are computed on the lattice of the finite
discrete Fourier coefficients.

The type and number of constraints is left free but must be linearly
obtained from the density field $\delta$. For example, one may put
constraints on the local shape of the density field and its peculiar
velocity at the same time. In this case, we have thus 10 constraints
for the local shape of the density field, which corresponds to its
value, its first and second derivatives. Additionally, we
have 3 constraints for the peculiar velocity. Once the constraints have
been put on the density, the rest of the density field is generated to
match a given cosmological power spectrum such that the statistics of
the final initial conditions are statistically correct.  We note that
this constrained realisation algorithm is fundamentally in Lagrangian
coordinates to set up initial conditions for the dynamics of
matter particles. This is particularly true for the choice of the size
of the filter in the kernel $K_i$, which must be understood in terms
of Lagrangian coordinates.

We implemented such an algorithm in a MPI parallel
environment.\footnote{This code is available publicly as ICgen on
 \url{http://www.iap.fr/users/lavaux/icgen/icgen.php}.} This code is capable of
handling an arbitrary number of Gaussian constraints of any size. The
memory and processor load is shared evenly among all nodes of the
cluster, making this code scalable on a MPI cluster of computer.

\subsection{The MAK reconstruction}
\label{sec:mak}

We propose to use the results obtained through the
Monge-Amp\`ere-Kantorovitch (MAK) reconstruction to find the
constraints to put on the initial conditions of an $N$-body
simulation. The MAK reconstruction is a non-linear scheme of
recovering the peculiar velocity field from an evolved non-linear
density field. It assumes that the dynamics has not suffered shell
crossing. In that case, it has been shown \citep{Brenier2002} that it is
possible to uniquely reconstruct the trajectories of the particles
from initial to present time.  Practically, the reconstruction
corresponds to finding the mapping $\sigma$ which minimises the
quantity \citep{Brenier2002}
\begin{equation}
  S_\sigma = \sum_{i=1}^N ({\bf q}_{\sigma(i)} - {\bf x}_i)^2,
\end{equation}
where the $\{ {\bf x}_i \}$ are the Eulerian coordinates of equal mass
particles composing the present day non-linear density field, $\{ {\bf
  q}_j \}$ are the homogeneously distributed Lagrangian coordinates of
this same distribution.  This method is thus purely
Lagrangian. Additionally, full mass conservation is hard encoded in
the method, removing the singularities that are present in linear
theory in the neighbourhood of large mass concentration.  The
assumption that no-shell crossing occurs is of course limited to
larger scales, typically a few megaparsecs, where structures are still in the
Lagrangian perturbative regime, called the Zel'dovich
approximation. This modelling of the dynamics is sufficient for our
purpose which is to compute initial conditions compatible with 
observations on large scales.

To handle redshift space distortions, we use the Zel'dovich
approximation to predict the peculiar velocities of particles at the
moment of the reconstruction.
The modified action has been shown to work well on large scales for the
reconstruction of peculiar velocities \citep{mohtu2005,LavauxTully09}.

This method is particularly well suited to work in conjunction with
the Hoffman-Ribak method as we derive all quantities in Lagrangian
coordinates.

\section{Computing initial conditions}
\label{sec:ic}

In this section, we describe the methodology used to go from the
original 2MRS data-set to the finished initial conditions of the
$N$-body simulation. Section~\ref{sec:data} describes the
data-set. Section~\ref{sec:constraints} describes how the constraints
were applied to the random Gaussian field. At the end,
Section~\ref{sec:icsetup} gives the details of the setup of the
initial conditions, in particular the choice of the background
cosmology.

\subsection{Dataset}
\label{sec:data}

To get better constraints for the simulation of the Local
Universe, we have used the 2MASS redshift survey, which has the advantage
of being full-sky, compared to other surveys, while being complete up
to $\sim$60-80\Mpch{}. To derive the mass distribution for the orbit
reconstruction, we assume a constant Mass-to-Light ratio. This
hypothesis has already given successful results on the comparison
between reconstructed and observed peculiar velocities using this
catalogue \citep{LavauxTully09}.

We use the mass splitter algorithm given in Appendix~A of
\cite{Lavaux08}. The algorithm looks for the optimal equal mass
splitting given that we have a fixed number of mass elements and that
the relative error between the splitted and the true mass must be as
low as possible. After the reconstruction of the displacements of the
equal mass particles, the global displacement of each of the groups of
galaxies is computed by a simple average of the displacements of the
particles representing this group.

 To mitigate boundary effects in the
inner part of the reconstruction volume, we use a 100\Mpch{} deep
reconstruction, padded with an homogeneous distribution of particles,
as specified in \cite{Lavaux08}, and have
extracted the objects within 60\Mpch{} for the Local Group. 

We have not used all of the objects for constraining the Gaussian
realisation as it would have yielded a 19913$\times$19913 matrix to
compute and invert. This would be expensive to compute exactly, and
additionally inversion would probably need to be regularised. Instead, we limit
ourselves to constrain the reconstructed peculiar velocities of the
bigger objects. We now describe the methodology used to constrain the
Gaussian realisation.

\subsection{Constraining initial conditions}
\label{sec:constraints}

We constrain the realisation within 60\Mpch{} of the centre of the
box, which corresponds to our mock observer, using the reconstructed
velocities obtained from the MAK reconstruction.  This reconstruction
has been shown to recover with precision the linear
regime from an evolved non-linear density field
\citep{Mohayaee05,Lavaux08}. We use the reconstructed 3D displacement
field in Lagrangian coordinates to constrain the motion of matter. We
choose an isotropic Gaussian smoothing window whose radius, $R_i$, is
determined by the mass of the considered object $i$ in the catalogue and
is given by:
\begin{equation}
  R_i = \frac{1}{\sqrt{5}}\left(\frac{3 M_i}{4 \pi \bar{\rho}}\right)^{1/3}.
\end{equation}
We add a $1/\sqrt{5}$ to match the Gaussian filter scale to a top hat
filter scale to second order in Fourier space. The top hat filter is a
good indicator of the filtering scale as the peculiar velocities are
computed by doing a simple average of all displacements of the
particles of a given group of galaxies when computing the total
displacement of this group. We neglect the impact of the actual shape
of the Lagrangian patch of the considered objects. To avoid having too
many constraints, we enforce that the constraint should at least have
a Lagrangian radius $R_i \ge 1.5$\Mpch{}. This corresponds to
considering only the peculiar velocities of groups of a mass greater
than $\sim 4\times 10^{12}$~M$_\odot$. This allows us to reduce the
size of the matrix to 3942$\times$3942. In doing so we make use of the
sensitivity of the peculiar velocities to large scale power in order
to reduce the computational complexity while keeping the results
intact. Indeed, for a typical \LCDM{} cosmology, the
  velocity field is correlated at $\sim90\%$ for distances less than
  $\sim 8$\Mpch{}. Consequently, we do not add much information by
  increasing the number density of constraints. Instead, we may even
  decrease the numerical stability of the method, which relies on a
  large matrix inversion. We discuss the impact of keeping only one
  fourth of these constraints at the end of
  Section~\ref{sec:dfield}. We expect that increasing the density of
  constraints should not change our results significantly. We include
  in Appendix~\ref{app:velocity_field_simu} a comparison between the
  simulated velocity field and the original constraints.

\subsection{Initial condition setup}
\label{sec:icsetup}

We use a \cite{BBKS} power spectrum with $n_s=1$,
$\Omega_\text{m}=0.258$, $\Omega_\Lambda = 0.742$, $H=72$~\kmsMpc, $\sigma_8=0.77$.  The time complexity increases
with the number of constraints $N_\text{C}$ as $O(N_\text{C} \times
N_\text{g})$, with $N_\text{g}$ the mesh size for generating initial
conditions. We choose a mesh of $N_\text{g}=256^3$ particles, sampling
a box of side $L=200$\Mpch{} for generating the constrained initial
conditions. This allows us to shape the white noise at a resolution of
$\sim$1\Mpch{} while keeping the time complexity low. At the end, we constrain 1,314 positions in a $\sim$60\Mpch{} volume in Lagrangian coordinates, which corresponds to 3,942 constraints. The computation
of the correlation matrix of the constraints at this resolution took a
little less than a week on 128 processors on the Mercury cluster at
NCSA. 
\section{Results}
\label{sec:results}

\begin{figure*}
  \includegraphics[width=\hsize]{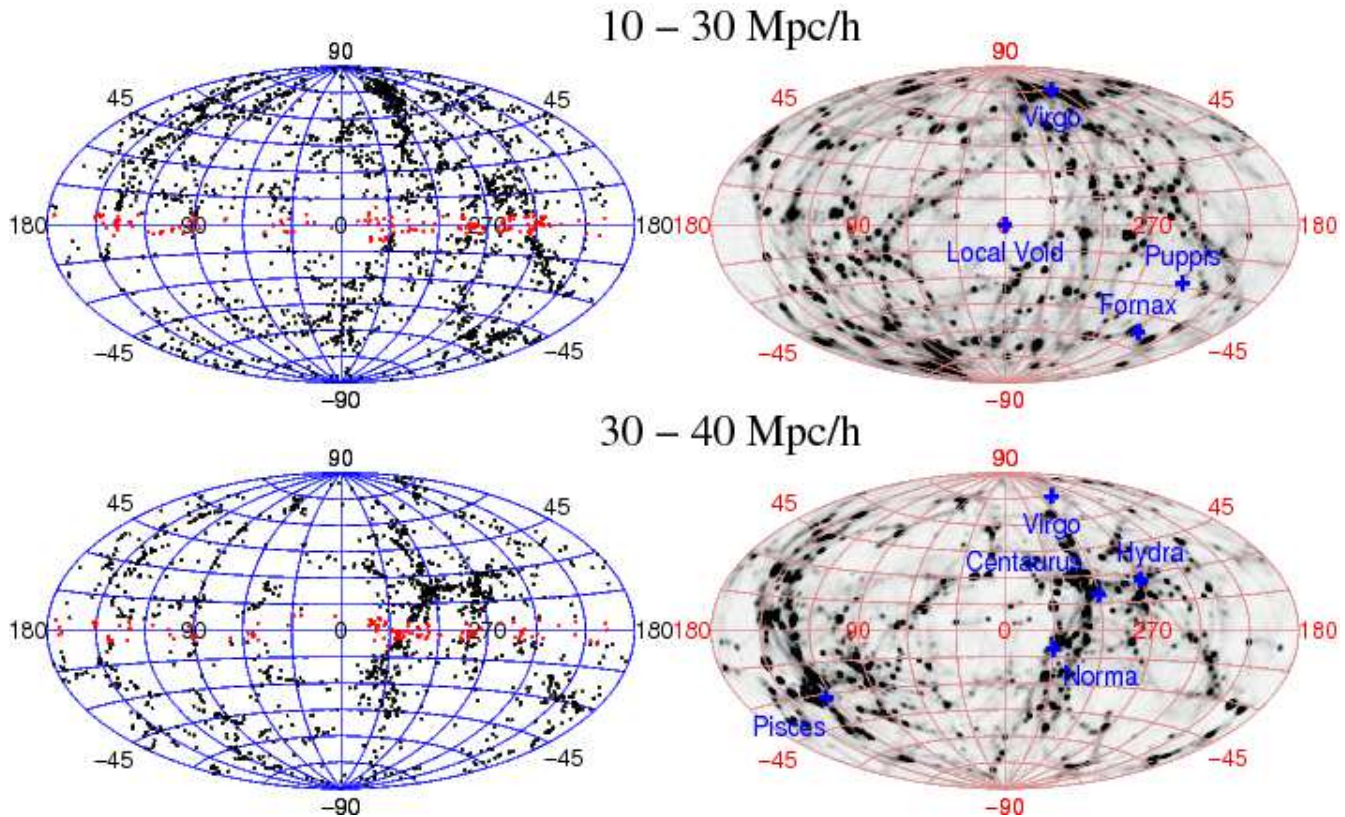}
  \caption{\label{fig:skymaps} {\it Observed vs simulated sky-maps} --
    Left column: Aitoff projection of the galaxies of the
    Two-Micron-All-Sky Redshift survey in different redshift slices,
    as indicated above the panel. Right column: Aitoff projected mass
    density in different redshift slices of the simulation, centred
    on the mock observer at the centre of the simulation. The contrast
    of the different slices have been adjusted for the distance of each
    slices.}
\end{figure*}

\addtocounter{figure}{-1}
\begin{figure*}
  \includegraphics[width=\hsize]{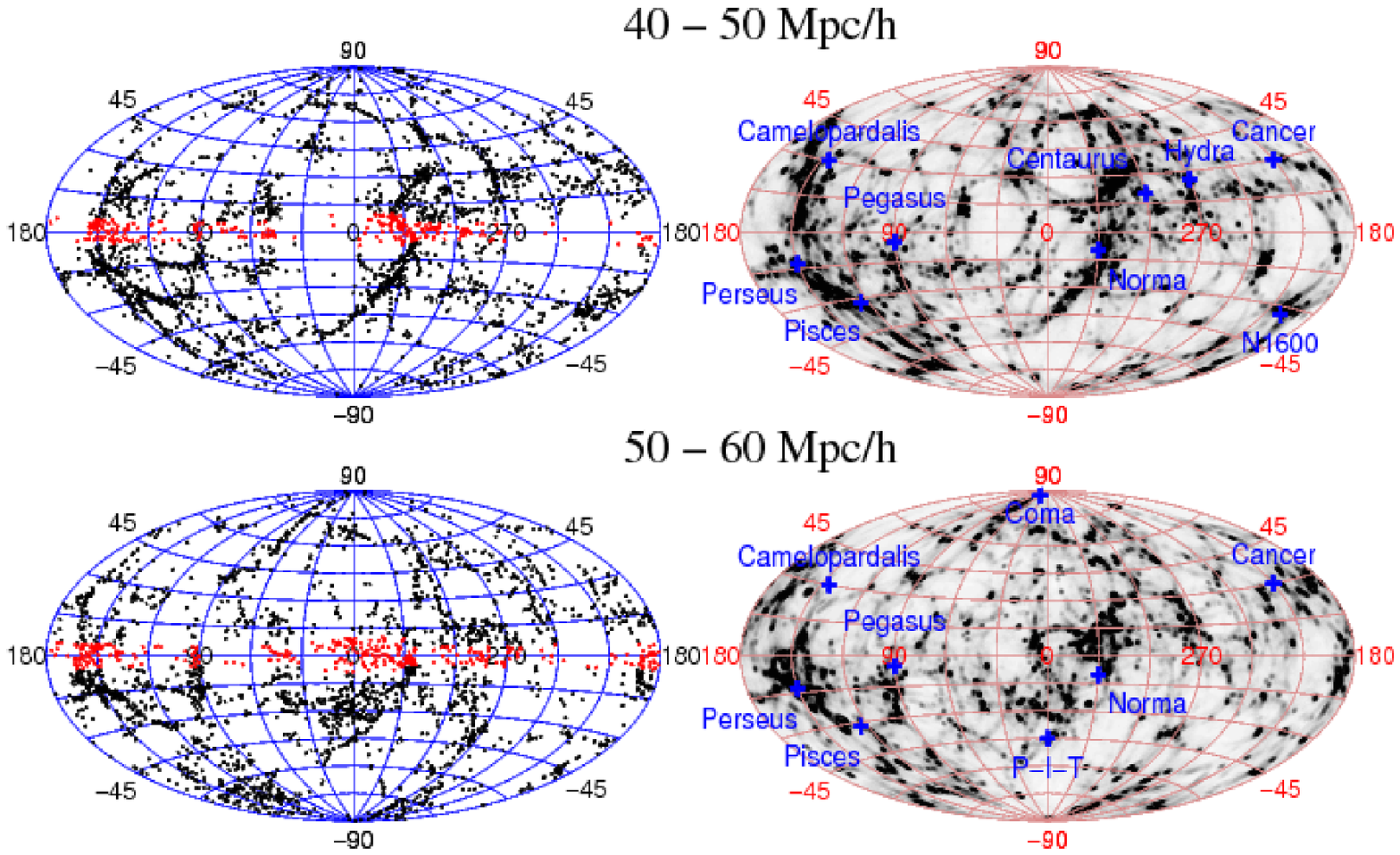}
  \caption{continued}
\end{figure*}

In this section, we detail the results of the constrained
simulation. We used the $N$-body simulation code
\verb,RAMSES,  \citep{ramses} to evolve the initial
  condition that are specified in Section~\ref{sec:icsetup}. The
  background metric is chosen to be a standard \LCDM{} cosmology with
  parameters $\Omega_\text{m}=0.258$, $\Omega_\Lambda=0.742$,
  $H_0=72$~\kmsMpc. We limit the number of refinement of the AMR grid
  to seven, this corresponds to soften the gravitational field on the
  scale of $\sim 6 h^{-1}$kpc, which is the size of one element of the mesh.
We first look qualitatively at the sky-maps of the distribution of the
dark matter. We then check the adequacy of the simulated
  velocity field to the observed velocity field within a volume of
  3,000~\kms{}. Finally, we compare quantitatively the simulated
  density field to the observed galaxy density field: by direct
  comparison and through the two point correlation function.

\subsection{Sky-maps}
\label{sec:skymaps}

In the left column of Figure~\ref{fig:skymaps}, we represent the
position of galaxies by redshift slices on an Aitoff projection of the
real sky. In the right column, we represent 
 the projected mass density of the simulation in these
same slices. The colour contrast is corrected such that the maximum
value represented in these sky map is proportional to $1/2 (d_{i+1} -
d_{i})(d_{i}+d_{i+1})^2$. This ensures that clusters are not
excessively dominating the map at larger distances compared to maps of
 the structures at smaller distances.

The visual comparison of the structures seen in the 2MASS
  Redshift Survey and in the simulation looks qualitatively in agreement. We reproduce nearly all the filamentary structure of the
Local Universe. We now look at the details of the structures in each of
the slices.

In the $10-30$\Mpch{} slice, Virgo is at its right place near the
North Galactic Pole. At the south pole, there is a significant dark
matter concentration that is probably linked to a set of galaxies
present in the 2MRS in the same slice. We correctly recover the Fornax
cluster at $(l,b)\simeq (236,53)$ with the Puppis cluster. We see that
the Local Void area is totally empty as it is expected from recent
peculiar velocity analysis \citep{Tully08}. This void unfortunately
corresponds to a region that is currently obscured by the galactic
bulge of our galaxy. It is present in the reconstructed velocity field
because we assume that the galaxies above and below the galactic
bulge were a fair representation of what is happening in the
direction of the bulge. As the number of galaxies is very low, it
happens that we have introduced a Local Void of the right size, as
expected from peculiar velocities. This void now appears in the
simulation.  Additionally, we recover all the observed filaments.  In
the right panel, we see the first tip of the Pisces cluster at
$(l,b)\sim (150,-30)$. We note a filament that goes from this cluster
to the Great Attractor region in the Hercules sector.  This filament
is lightly visible in the distribution of galaxies on the left panel.

In the $30-40$\Mpch{} slice, the galaxy distribution features are
almost completely reproduced in the dark matter density field of the
$N$-body simulation.  The first tip of the Pisces cluster appear in
that slice, even though it should be mostly placed in the the
50\Mpch{} region.  In the Great Attractor direction, $(l,b)\sim
(300,15)$, we see that the simulation follows very precisely the
distribution of galaxies on the left. The two concentrations of
galaxies in the middle of the sky projection correspond to the
Hydra-Centaurus supercluster, as highlighted in the right panel. The
filaments are correctly visually interconnected according to the
galaxy distribution. The Local Void is still present in this slice.

So far, the number of galaxies introduced to fill the Zone of
Avoidance was small. In the $40-50$\Mpch{} slice, this number
increases. We recover all marked structures of the right panel of this
slice. We also recover the structures introduced by the filling of the
Zone of Avoidance. For example, the ``S'' shape in the direction
$(l,b)\sim(300,0)$ is recovered in the distribution of dark
matter. So, the constrained initial condition are working even better
than expected in that respect, even though this shape may be an
artificial construction due to the Zone-of-Avoidance filling using mirrored galaxies.

The galaxy distribution in the $50-60$\Mpch{} slice is totally reproduced in 
the dark matter density field of the simulation.

\subsection{Velocity fields}
\label{sec:vfield}

\begin{figure}
  \includegraphics[width=\hsize]{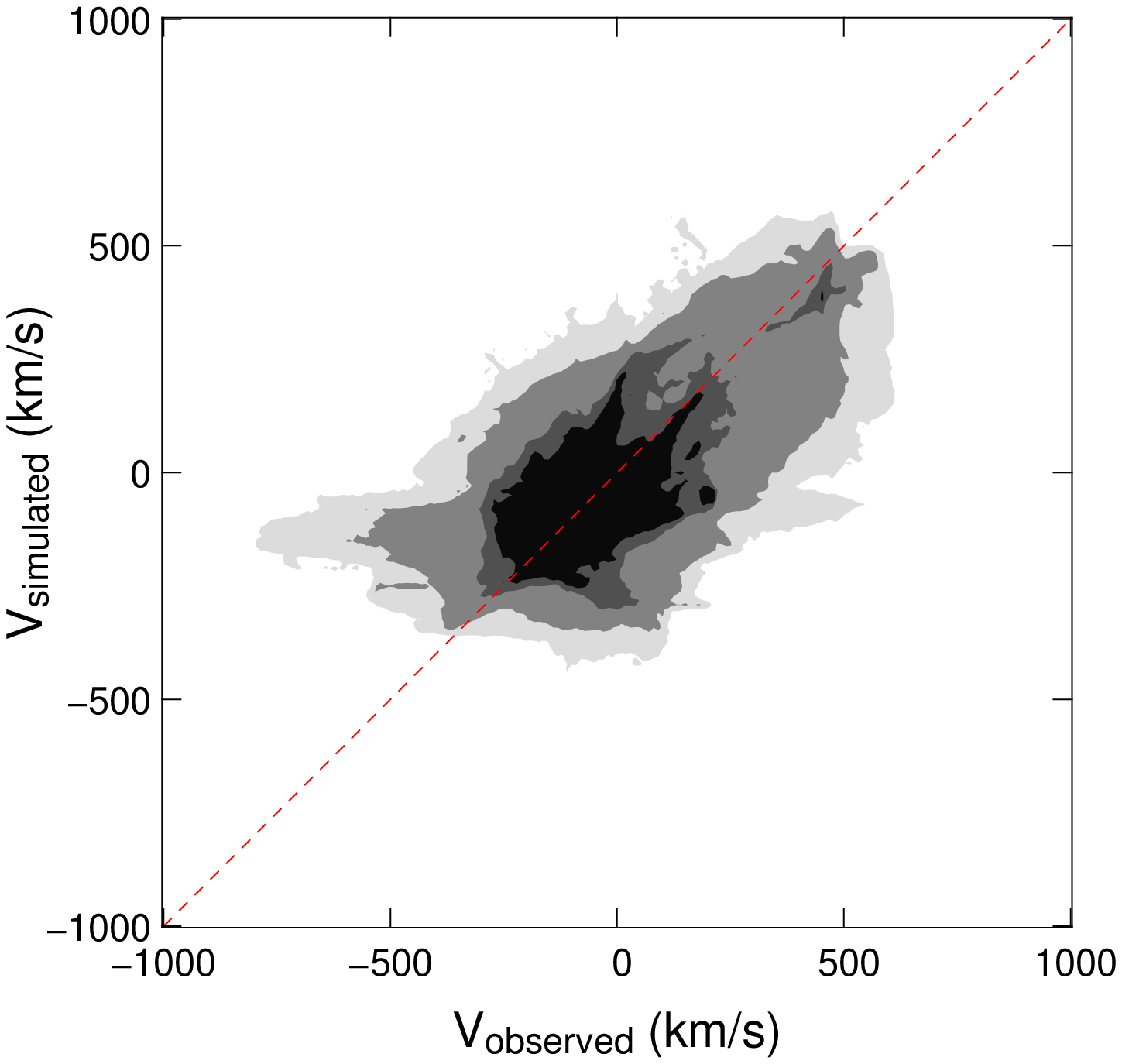}
  \caption{\label{fig:vfield_obsrec} {\it Simulated peculiar
      velocities along the line-of-sight vs. observed peculiar
      velocities} -- We represent here the direct comparison of the
    simulated velocity field against the observed velocity field
    (line-of-sight component) within the 30\Mpch{} volume. We
    subtracted the 30\Mpch{} bulk flow in the two cases. The filled
    contours corresponds to 50\% (black), 68\% (dark grey), 95\%
    (grey), 99\% (light grey) of the 30\Mpch{} volume. Both velocity
    fields were adaptively smoothed with a spline of radius in the
    range $[2.5;13]$\Mpch, with a median of 7\Mpch{}.}
\end{figure}

In Figure~\ref{fig:vfield_obsrec}, we compare quantitatively the
simulated line-of-sight component velocity field to the observed
line-of-sight component of the velocity field.  For obtaining the
observed velocity field we used the extended NBG-3k catalog
\citep{Tully08}. The distance were matched to the group of galaxies of
the 2MASS Redshift Survey galaxies so that 726 groups receive a
distance \citep{Crook07,LavauxTully09}. We did an equivalent
treatment for the simulation. We first grouped the particles
according to a Friend-Of-Friend algorithm with a linking length of
$0.2\bar{n}^{-1/3}$, with $\bar{n}$ the mean particle density of the
simulation \citep{E88}. We applied a threshold of having a minimum of
8 particles. This grouping helps at reducing
Finger-Of-God phenomena while keeping other redshift distortion
effects.

The two velocity fields are computed in
redshift space to avoid systematic effects \citep[e.g.][]{LB88,DBF90}.
Moreover, to avoid being contaminated by
Finger-of-god effects, we first computed groups using a
Friend-Of-Friend algorithm in the simulation (linking length equal to
0.2 the mean inter-particle length). We also used the distance to
grouped galaxies obtained from the anisotropic Friend-Of-Friend that we used
to make the groups of galaxies in the 2MASS Redshift Survey. This anisotropic Friend-of-Friend is described in \cite{HG82,Crook07,LavauxTully09}.

The two velocity fields agree well with each other, although there is a
large scatter. This scatter has at least three origins: the modelling
error of the dynamics in the reconstruction of the orbits of the
galaxies of the 2MRS, the modelling error of the Zel'dovich transport
when applying the constraints coming from reconstruction and the
measurement errors on the observed velocity field.

\subsection{Density field}
\label{sec:dfield}

\begin{figure}
  \includegraphics[width=\hsize]{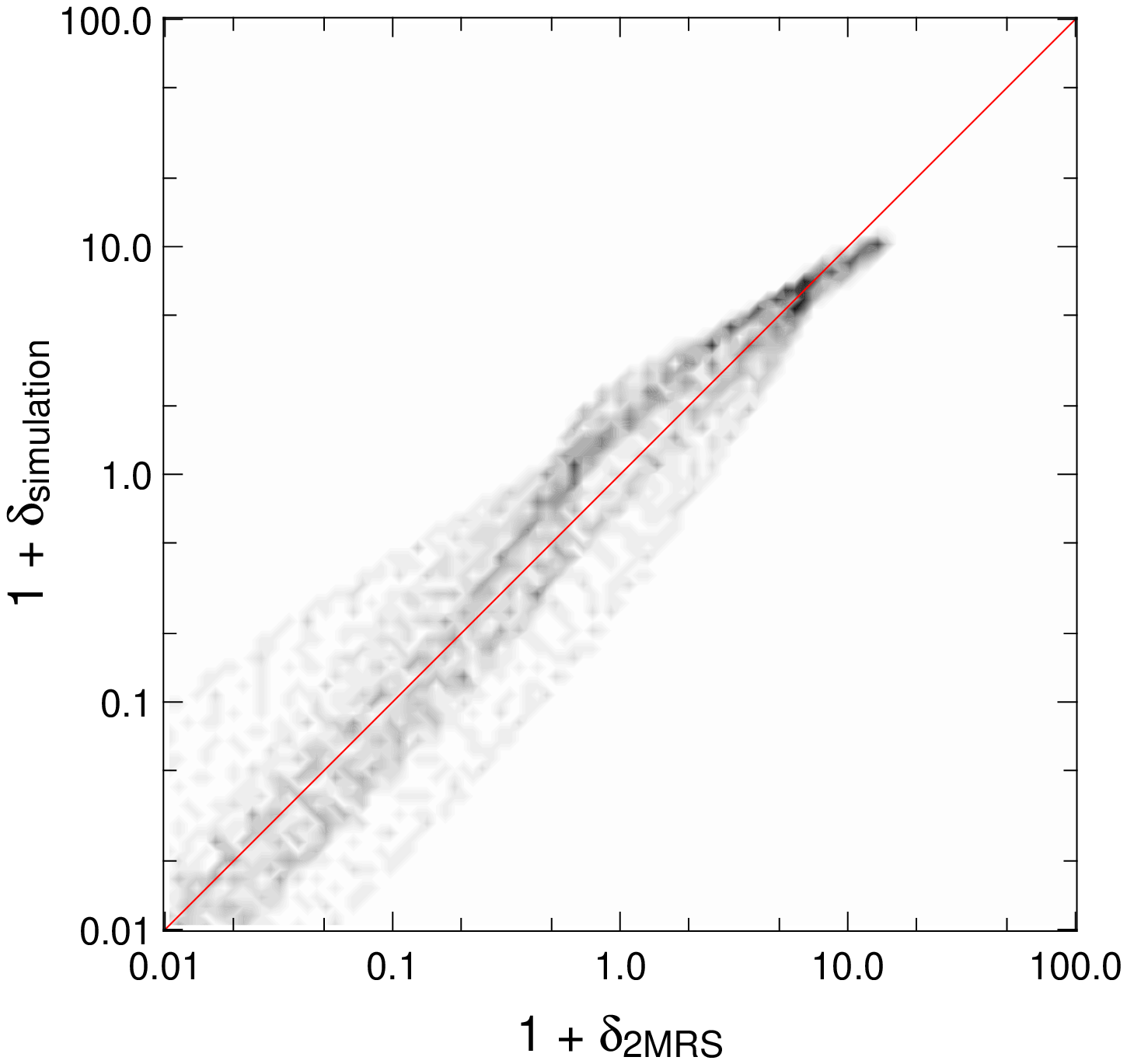}
  \caption{\label{fig:den_2mrs_simu} {\it Correlation of the density
      field of the simulation to the density field of the 2MRS
      catalogue} -- We have extracted a volume limited sample of
    60\Mpch{} from the 2MRS and the equivalent volume from the
    simulation. The density fields are both smoothed to 3.5\Mpch{}
    with a Gaussian kernel.}
\end{figure}

In Fig.~\ref{fig:den_2mrs_simu}, we show the correlation of the
density field of the simulation and of the 2MRS galaxy number,
smoothed to 3.5\Mpch{} with a Gaussian kernel. We define the correlation coefficient
as the ratio:
\begin{equation}
  r = \frac{\langle \log(\rho_\text{sim}/\bar{\rho_\text{sim}}) \log(\rho_\text{obs}/\bar{\rho_\text{obs}})\rangle}{\sqrt{\langle (\text{log}(\rho_\text{sim}/\bar{\rho_\text{sim}}))^2\rangle \langle (\text{log}(\rho_\text{obs}/\bar{\rho_\text{obs}}))^2 \rangle}},
\end{equation}
with $\rho_\text{sim}$ the smoothed density field of the simulation, $\rho_\text{obs}$ the smoothed galaxy number density field in a volume limited sample 2MASS Redshift Survey.
 The correlation
coefficient of the logarithm of the density field is 0.97.
This correlation is not perfect. On the high end, we see that
the main alignment of the scatter distribution is not strictly
following the red diagonal. This could be an effect of the galaxy bias
at that scale. 
 Reducing the number of constraints on the
   initial condition from 3,942 to $\sim$1,000, either by selecting
   more massive objects or by randomly picking up one constraint of
   four, leads to a decrease of the correlation of the two density fields to $0.92$. Additionally, we see a significant increase in the scatter
in the high density region, the low density being apparently unaffected. The scatter in the simulated velocity field remains unaffected by this change. 

\subsection{Two point correlation function}
\label{sec:2pt}

\begin{figure}
  \includegraphics[width=\hsize]{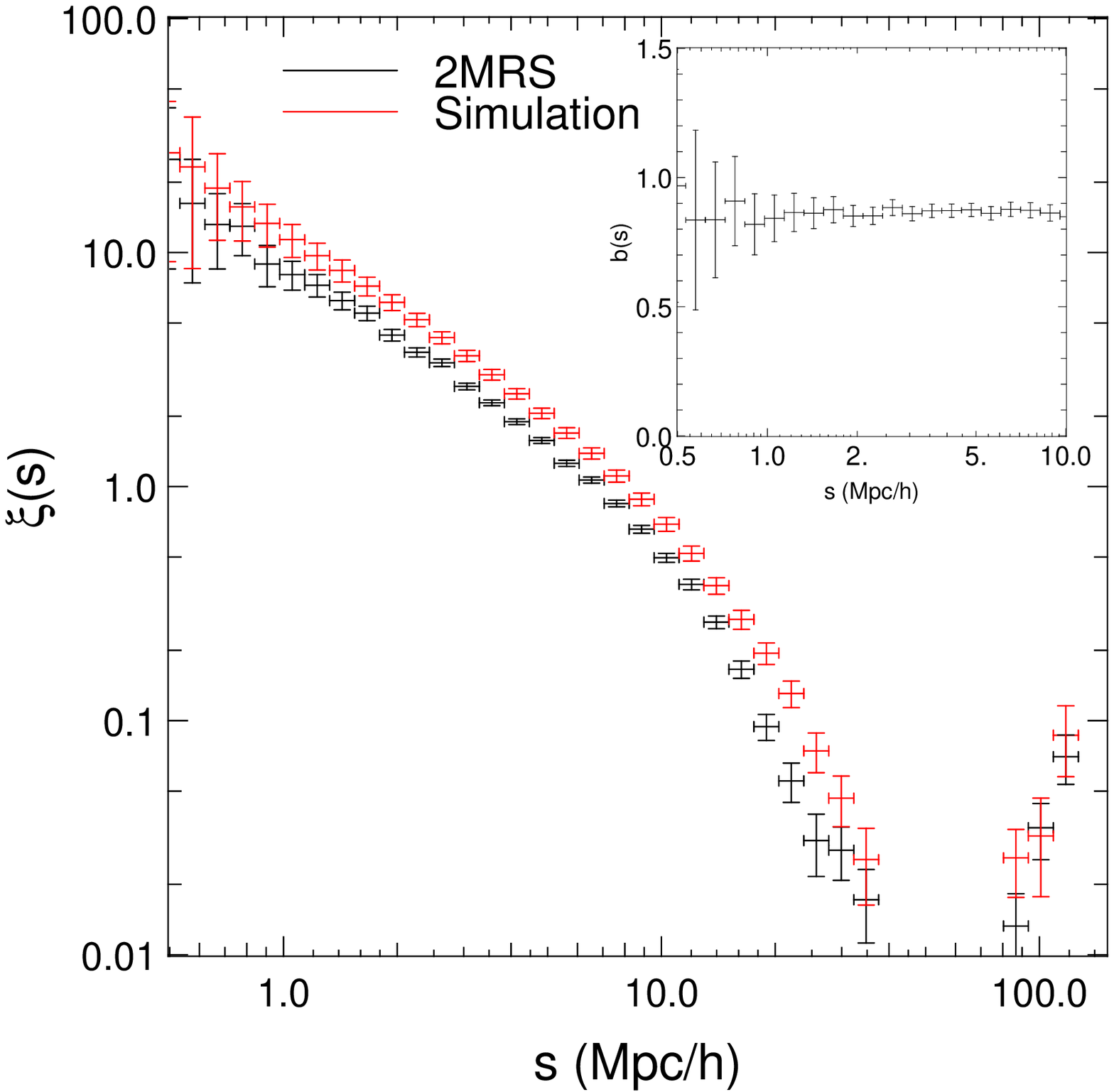}
  \caption{\label{fig:2pt} {\it Two point correlation function of the
      galaxies and of the dark matter in the simulation} -- We have
    computed the two-point correlation function of the dark matter
    particles of the simulation (red) and of the galaxies of the 2MASS
    Redshift Survey (black) for the volume limited sample of 60\Mpch{}
    of volume. We represented the ratio $b^2 =
    \xi_\text{2MRS}/\xi_\text{simulation}$ in the inset plot. The
    error bars are estimated using a Jackknife test on 1,000 sub-volumes 
    with 1,000 different realisations of the sampling noise of the volume.}
\end{figure}

Another quantitative comparison between observations and simulation is
given by the two point correlation function. Comparing the two point
functions given by the dark matter distribution of the re-simulation of
our Local Universe and the actual two point function of the galaxies
of our Universe may allow to have a better understanding of the galaxy
distribution bias by direct measurement. We propose here a simple test
to estimate the scale dependent bias in our specific part of the
Universe by taking the ratio of these two functions.

We use the estimator of \cite{LS93} to compute the correlation function
within a 60\Mpch{} volume:
\begin{equation}
  \xi(s) = \frac{DD(s) - 2 DR(s)+RR(s)}{RR(s)},
\end{equation}
with $s$ the amplitude of the redshift separation.
We have built a volume limited sample of the 2MRS such that it is
complete up to a distance of 60\Mpch{} from the observer, which
corresponds to keeping 5,472 galaxies.  In this test, we are not
sensitive to cosmic variance effects as we are looking at the same
part of the Universe both in the simulation and in the observations. So,
the error bars on our correlation function should only reflect the
sampling errors.  To estimate these errors, we have sub-divided the
60\Mpch{} volume into 1,000 sub-volumes and performed a Jackknife test
by computing the correlation function on the whole 60\Mpch{} except
the concerned volume. For each of the computation of the correlation
function, we have used a different realisation of the sampling noise
used to compute $RR$ and $DR$.  This generated 1,000 correlation
functions compatible with the data.

We have computed the correlation function of the simulation in exactly
the same way. We have first extracted randomly 5,472 particles, and we
have applied the same cuts in volume as for the catalogue, including
the Zone-of-Avoidance effect.

The results are given in Fig.~\ref{fig:2pt}. The central point of the
correlation function corresponds to the average of all the obtained
correlation function by the Jackknife test and the Poisson sampling
test. The vertical error bars reflect the standard deviation of
$\xi(s)$ according to its average. The horizontal error bar gives the
size of the bin.

 We see that the two correlation function are closely following each
 other except on the smaller scales. This is expected as we implicitly
 assumed in the reconstruction that galaxies are fair tracers on large
 scales.  We note that, in our case, the value of the bias for the
 galaxies of the 2MRS seems to be $b=0.87 \pm 0.03$. However, this
 result is probably systematically affected by the particular value of
 $\sigma_8$ for the realisation of the large scale mode of the
 simulation, which are more compatible with $\sigma_8=0.83$ instead of
 $\sigma_8=0.77$. This can introduce a small bias of the order of
 $0.93$, which corresponds to what we observe here.  Additionally, the
 fair comparison between the simulated velocity field and the observed
 velocity field (Fig.~\ref{fig:vfield_obsrec}) reinforces this result
 as the velocity field is sensitive to the total matter distribution,
 even though we used a constant mass-to-luminosity ratio to derive the mass of each galaxy to reconstruct their orbits.

\section{Conclusion}
\label{sec:conclu}

We have implemented a new method to re-simulate the Local Universe
with a standard $N$-body simulation code from the 2MRS catalogue of
galaxies. This method allows
us to reproduce in many details the Large-scale structures of the
Local Universe up to 60\Mpch{} and to a precision of a few
megaparsecs. This method is easily applicable to larger volumes and
surveys. We illustrate the quality of the simulated volume by direct
comparison within a 30\Mpch{} volume of the simulated velocity field
to the observed velocity field. Furthermore, we  highlight a
direct application of this simulation by the joint measurement of the
two point correlation function in the simulation and in the actual
observations of the positions of the galaxies of the 2MRS.

This method opens us the possibility of studying quantitatively both larger and
smaller scales for our Local Universe. On small scales,
such a simulation would allow to run a simulation of the Local Group
including environmental effects. On large scales, we would like to
extend this work to the quantitative comparison between CMB data and
Large Scale structure data, with the major advantage, compared to
earlier attempts, of taking into account non-linear dynamics and
clustering. Along with semi-analytic modelling of galaxy formation, it
is possible to build a non-statistical simulation of the galaxies in our sky, which would enable us to
make a direct quantitative comparison with observations.

\section*{Acknowledgements}

The author thanks S. Colombi, R. Mohayaee, C. Pichon, B.~D. Wandelt \&
R.~B. Tully for useful discussions and suggestions.  The author
acknowledge financial support from NSF Grant AST 07-8849. This
research was supported in part by the National Science Foundation
through TeraGrid resources provided by the NCSA under grant number [TG-MCA04N015]. Teragrid systems are
hosted by Indiana University, LONI, NCAR, NCSA, NICS, ORNL, PSC,
Purdue University, SDSC, TACC and UC/ANL. The author thanks the French ANR
(OTARIE) for support.

\clearpage
\appendix
\section{Comparison of the reconstructed to the simulated velocity field}
\label{app:velocity_field_simu}

\begin{figure}
  \includegraphics[width=\hsize]{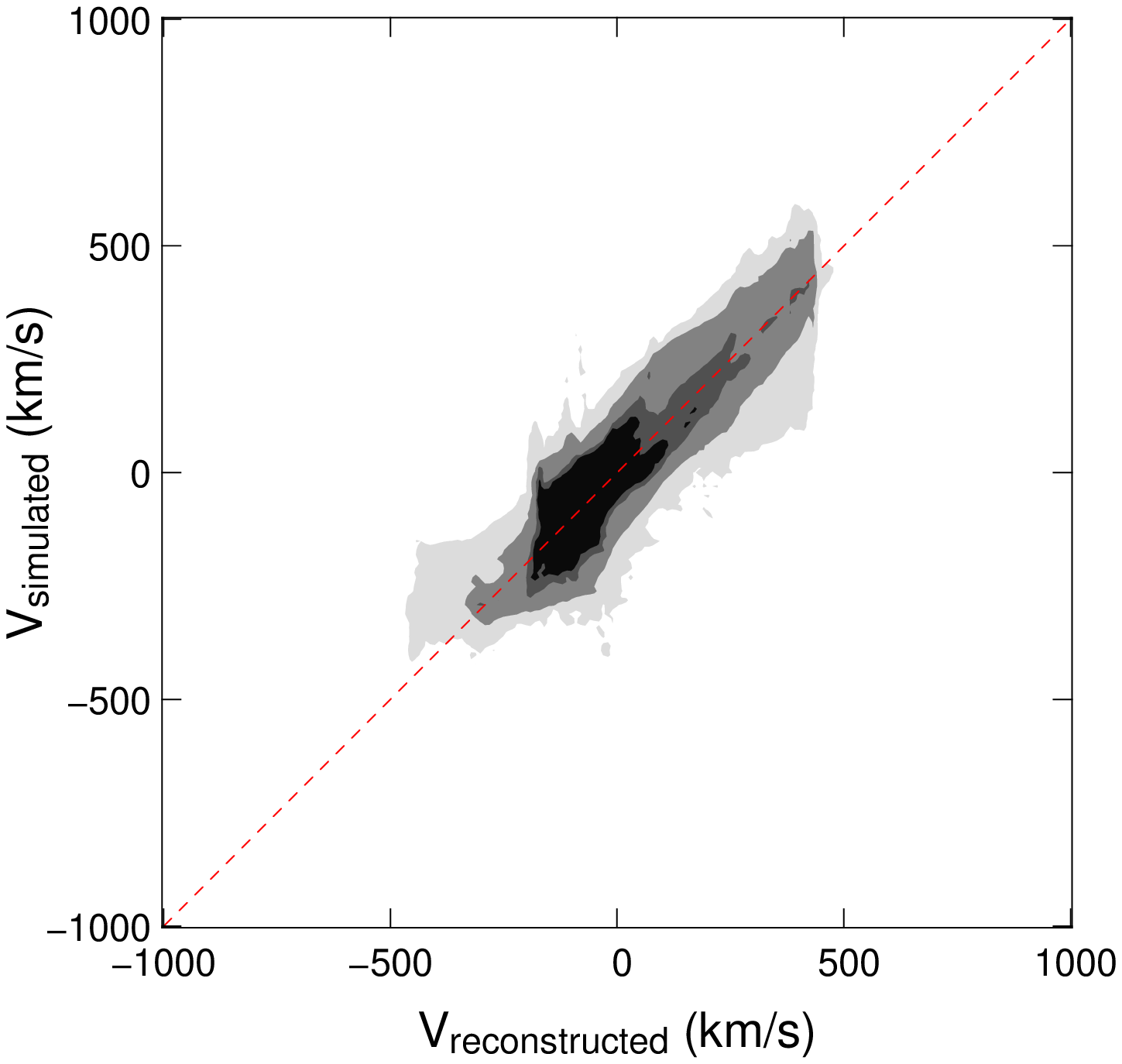}
  \caption{\label{fig:sim_rec_field} {\it Simulated velocity field
      vs. original reconstructed velocity field} -- We represent here
    a quantitative comparison of the originally reconstructed velocity
    field, which was used to constrain the initial conditions, against
    the final simulated velocity field. We used the same colour convention and the same smoothing procedure as
    in Figure~\ref{fig:vfield_obsrec}. }
\end{figure}

We represent in Figure~\ref{fig:sim_rec_field} a comparison of the
original reconstructed velocity field, used to derive the constraints
on the initial conditions, to the simulated velocity field, obtained
by evolving the initial condition using an $N$-body simulation. We
note that the two fields are fairly equal up to some scatter that is
probably due to small scale clustering, which is not modelled by the
MAK reconstruction, and the imperfections of the Zel'dovich
approximation, used to derive the velocity field constraints.

\end{document}